# Using Genetic Algorithms to Benchmark the Cloud

Jeff Kinnison and Sekou L. Remy

School of Computing

Clemson University

Clemson, South Carolina 29634

Email: {jkinnis, sremy}@clemson.edu

**Abstract**

This paper presents a novel application of Genetic Algorithms(GAs) to quantify the performance of Platform as a Service (PaaS), a cloud service model that plays a critical role in both industry and academia. While Cloud benchmarks are not new, in this novel concept, the authors use a GA to take advantage of the elasticity in Cloud services in a graceful manner that was not previously possible. Using Google App Engine, Heroku, and Python Anywhere with three distinct classes of client computers running our GA codebase, we quantified the completion time for application of the GA to search for the parameters of controllers for dynamical systems. Our results show statistically significant differences in PaaS performance by vendor, and also that the performance of the PaaS performance is dependent upon the client that uses it. Results also show the effectiveness of our GA in determining the level of service of PaaS providers, and for determining if the level of service of one PaaS vendor is repeatable with another. Such a concept could then increase the appeal of PaaS Cloud services by making them more financially appealing.

# 1  Introduction

A wide range of cloud services are available for application development and hosting, and each service is unique due to differences in implementation. These differences manifest when running applications, thus performance assessments are necessary to determine the suitability of a particular service to the needs of an application. Assessments typically consist of benchmarks reporting running time, memory usage, disk read/write operations, or other relevant metrics.

This paper proposes a new kind of benchmark carried out using population-based search methods. Population-based search is well-suited for benchmarking cloud systems because parameters can easily be tuned to test the operating boundaries of both the application and cloud service. Moreover, these methods may be implemented across multiple types of computing systems because they are not constrained to a particular computing platform.

The population-based search method presented in this paper uses an implementation of a genetic algorithm to provide potential solutions to a cloud-hosted application. This implementation has two main benefits. The first is that it may be replicated easily; the second, that it is not bound to a particular operating environment. This genetic algorithm was used to benchmark the running time performance of a cloud-hosted application to demonstrate its utility.

# 2  Background

## 2.1  Benchmarking

Benchmarking is the strategy used to evaluate system performance. In modern computing systems, hardware, software, storage, and the network all have impact on the overall system performance. The configuration of the system, must include consideration of all these components must be considered when discussing a benchmarking strategy.

Cloud Computing is a computing paradigm that adds to the challenge of undertanding the current state of the system. There are different cloud service models and options that a particular software suite can be implemented upon. As an example, for Python programs, virtual environments like Venv, containers from vendors like Docker or Solaris, and Virtual Machines, are all options along just one dimension of changes that can impact performance.

As such, it has been argued that traditional benchmarks like TPC [1], are not sufficient for analyzing the novel cloud services because they do not characterize properties like scalability, pay-per-use, fault tolerance, and they also make it difficult to generate comparable results since services with different capabilities and guarantees could be utilized in the system [2]. This is not to say that benchmarks like TPC-W [3, 4] were not been developed, but their usefulness for those considering benchmarking the cloud for High Performance Computing applications is limited.

## 2.2  Cloud Benchmarking

Cloud technologies have been investigated for their ability to run high-performance computing applications [5, 6, 7, 8, 9, 10]. Current benchmark research has focused on characterizing individual vendor's services [11, 12, 13]. However, it has been limited to the Infrastructure as a Service (IaaS) model [14, 13, 15]. To obtain a better understanding of the state of cloud computing, investigations beyond individual IaaSs are needed.

A prominent alternative to IaaS is Platform as a Service (PaaS). Unlike IaaS, which gives users administrator access to a full virtual machine (VM), PaaS performs system administration behind-the-scenes and provides identical virtual machines to each user [16]. This cloud computing model trades precise environment management for ease-of-use, allowing entities with limited resources or system administration experience to run HPC applications. Despite this advantage, limited formal research, like [17, 18], has been conducted regarding the applicability of PaaS to this domain.

To address this issue, this project investigated the behavior of PaaS in support of a computationally intensive application in conjunction with local client machines. Specifically, it benchmarked the total running time of a population-based algorithm communicating with a PaaS-hosted fitness assessment. To enable a nuanced understanding of the service model, times to completion were collected using PaaS environments from three distinct vendors coupled with three distinct types of client machines. The data were collected for two purposes: to investigate the ability of a PaaS to run an HPC application and to shed light on behavioral trends across different PaaSs. In terms of Testing and Validation, there are recent publications like [19] that indicate the challenges of comprehensive testing in cloud environments. Their approaches provide useful insight on the nature of the issues that need to be studied, but as their goal is not performance assessment, they do not package information for developers in that manner. It should also be noted that while it it common to consider PaaS and IaaS as distinct service models, there are some products, like Azure or Openstack, that provide cloud offerings that do not fit cleanly into these categories. There are benchmarks suites like [20] that are being developed to provide information about this class of cloud offerings. While these resources are useful, they suffer from some of the same challenges as TPC in that the do not generate comparable benchmarking results.

The broader impact of this work is a framework that not only can be used for benchmarking, but it can also be used to advance both education and research in industry and academia. Specifically, by leveraging a common infrastructure deployed via the Cloud, a common set of optimization problems can be shared and the derived solutions pooled. Further, since the same problems are being solved in each case, the comparision of optimization algorithms can focus on their particular research contribution without having to be concerned with the development of the underlying infrastructure.

## 2.3 Genetic Algorithm Benchmark

Genetic Algorithms [21, 22] are a population based search method which can be used to optimize a set of parameters in a search space. The technique harnesses a large number of candidate solutions to explore the search space, and to exploit the information accrued during the search. This approach uses biologically inspired operations, appropriated from evolutionary biology, to modify the candidate solutions at each stage of the search (called a generation). The success of this adaptive approach in solving problems associated with a large search space has led to numerous applications of the technique to a wide range of disciplines [23].

The solutions selected for this work were the parameters for a controller for a simulation of a ball-on-plate system. This controller was defined by a set of six, real-valued parameters that were used to determine the control output to modify the orientation of the plate. By controlling the plate, the position of the ball which rolled upon it was influenced. This system, or plant, is well studied in control theory, and provides an effective canonical optimization problem. The fitness of a considered controller was derived by quantifying the error observed when that controller was set to follow a predefined trajectory. The smaller the error, the fitter the controller.

The process of evolving candidate solutions is comprised of six stages that are applied repeatedly (see Figure 1). The stages include genetic material creation, plant simulation, fitness assessment, crossover and mutation (and creation of new genetic material from an existing population). Since the simulation and fitness assessment for each member of the population, that are performed at during each generation of the evolution are independent of the calculations for the other members of the same generation, this sequence of calculations can be performed in parallel. Further, if enough computing resources exist, the entire populations' fitnesses can be calculated at the same time.

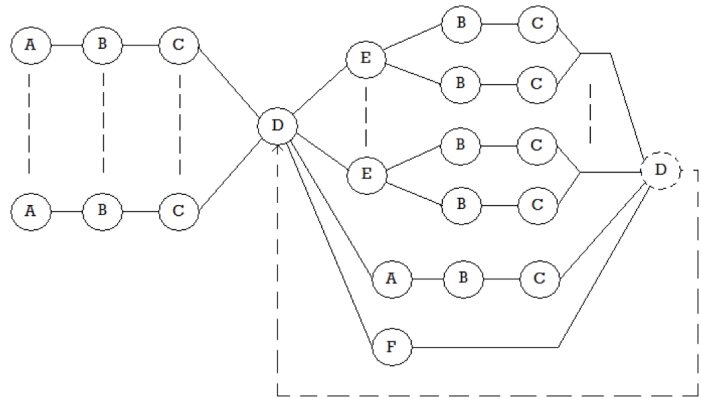

Figure 1: A: genetic material creation, B: Plant simulation, C: Fitness assessment, D: Collection and selection, E: Crossover and mutation, creating two candidate controllers. F is a dummy process that enables elitism. This is repeated for each generation of the population's evolution (dotted arrow).

In this project we implement the simulation and fitness assessment as a webservice. The client GA code thus makes HTTP requests to evaluate the fitness of each of the candidate solutions in the population. Each iteration, or generation, contains the same number of individuals created either from scratch or by altering the parameters of the members of the previous generation. The main parameters that determine a GA's computational workload, then, are the size of a population and the number of generations to evaluate.

The parameters of a GA can easily be altered to scale computational workload—the number of fitness evaluations computed—so it was used to stress-test individual PaaS environments. In this project, fitness values ranged from -99 to 400, and the seed solution generated a fitness score of 215. If the first generation of the GA did not contain at least one individual with a score of 215 or higher, then that run was considered invalid. Invalid populations were possible since the network and the PaaS service were both infrastructures that were not under the full control of the experimenters.

In addition to using a population seed, this GA implementation applied the concept of elitism. This concept guarantees that the best solutions from one generation also make it to the next. As a result, the fittest solutions were not lost in the course of creating new generations. Given that the GA communicated with a PaaS application over a network, these two additions to the general GA allowed for automatic error recovery while running this experiment and quick error detection afterward.

# 3 Experimental Setup

This experiment was conducted using both PaaS environments and local computers. The PaaS portion of the infrastructure was used to support an HPC application, and the local computers individually communicated with it. The combination of PaaS and client computation permits a scalable computational workload, used here to assess the performance of PaaS environments.

## 3.1 PaaS Implementations

To enable a comparison of PaaS environments, three were selected to host identical copies of the same application. The resources associated with each PaaS were limited to free-tier offerings, meaning that no application had access to more than the most basic services provided by its host. The most significant constraints on each PaaS are listed in Table 1. PaaS2 had the most stringent quota, so PaaS1 and PaaS3 were run within the same constraints as PaaS2 to unify testing. Staying within every quota was necessary because, once a quota was met, the PaaS shut down the application until the start of the next quota period.

Table 1: Constraints on Free Services for Each PaaS

| PaaS | Most Significant Constraint on Use |
|---|---|
| 1 | 720 uptime hours per month |
| 2 | 28 uptime hours per day over all active processing units |
| 3 | 512 MB storage for application code |

For this work, the optimization target was the set of parameters for a program that controlled the position of a ball on a smooth plate[24]. This problem is a classic problem from control theory, and was selected since it provides a challenge with multiple known solutions/approaches. The controller and the model of the system are both implemented within the same codebase, and due to the limitations of PaaS infrastructures, they are implemented as a discrete event simulation. The fitness of a particular set of parameters is defined as a function of the error from the target trajectory of the ball. The smaller the error, the larger the fitness, thus this problem is phrased as a maximization problem. Each of the three PaaS vendors permitted the use of the web.py framework[25], so the application's web service was implemented using this psuedo-standard resource.

## 3.2 Clients

The hardware specifications of each client computer are listed in Table 2. Client-A served as the control because it was a built to run HPC applications on a university campus, while Client-B was a workstation in a computer lab. These machines were chosen because they represent a important points on the spectrum of computing resources used for HPC.

Table 2: Client Hardware Specifications

| Hardware | Client-A | Client-B |
|---|---|---|
| CPU | x86_64 Xeon | x86_64 i7 |
| Processors | 24×2.40 GHz | 4×2.90 GHz |
| RAM | 96GB | 6GB |

## 3.3 PaaS/Client Interaction

Client and PaaS communication only occurred through HTTP requests, and every client accessed each PaaS for a total of nine PaaS/client pairs. Every client ran the same number of concurrent populations with a fixed population size. Each individual sent one HTTP request to the PaaS application per generation. Recording runtimes with respect to PaaS/client pairs enabled performance examination in terms of different clients communicating with the same PaaS and of the same client with different PaaSs, allowing trends to emerge from both perspectives.

The GA parameters were investigated first to determine a feasible population size and number of populations of equal size which could be run in parallel. As a consequence of each individual in a population concurrently accessing the PaaS infrastructure, large populations could result in most requests returning timeout errors, which would produce inaccurate runtime results. The study began with one population of size 125 and gradually decreased the number of individuals in a population while increasing the number of concurrent populations. This first phase determined that one trial could consist of 10 concurrent populations of 50 individuals running over 80 generations for every PaaS/client pair.

Once the GA parameters were set, each client was again tested in conjunction with each PaaS to determine the interval at which trials could be run without overlapping. This phase was carried out for the same reason as the first: overlapping trials could result in a majority of timeouts. While the viable testing interval varied between PaaS/client pairs, it was determined every pair could complete one trial per hour with no overlap, and four trials could be conducted per PaaS every day.

## 4 Experiment

Every trial in the data collection phase consisted of one client communicating with one PaaS using the GA parameters found in the pilot study. Before starting the GA, a timestamp was recorded. The GA ran to completion, and then a second timestamp was recorded and the input and fitness values of all individuals from every generation were saved to a file with the two timestamps for future analysis.

Trials were randomly selected and examined to monitor the status of the experiment. In addition, after all data were collected, every population was examined so that all failed results could be removed along with corresponding populations from every other PaaS/client pair. A population failed if no fitness value in the first generation exceeded one, indicating network difficulty during that trial. All population data that passed this test were analyzed.

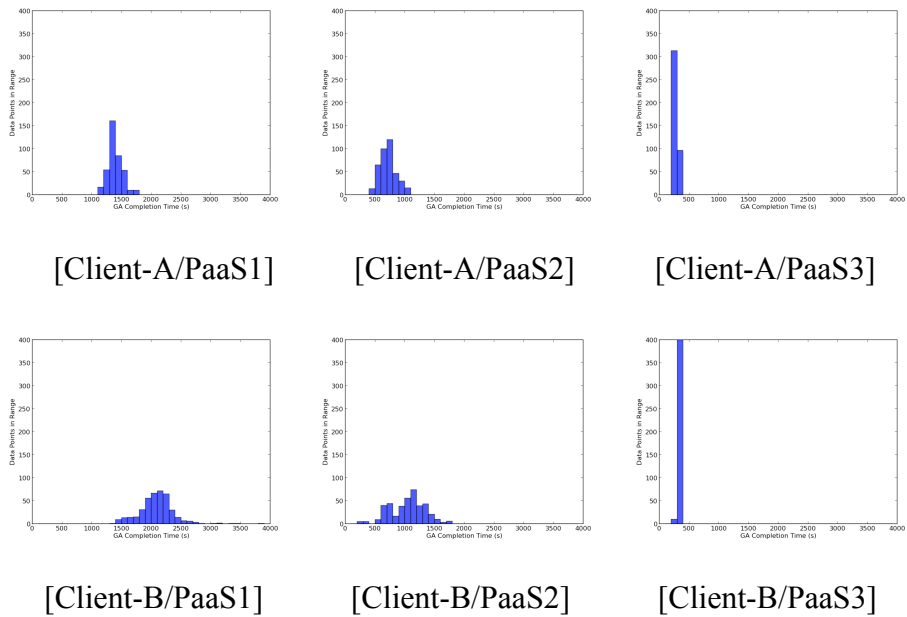

[Client-A/PaaS1]   [Client-A/PaaS2]   [Client-A/PaaS3]

[Client-B/PaaS1]   [Client-B/PaaS2]   [Client-B/PaaS3]

Figure 2: Runtime Distributions of PaaS/Client Pairs.

# 5 Results

Each PaaS/client pair underwent 41 trials, generating 410 runtimes per pair. The distributions of the runtimes are presented in Figure 1. This chart shows that there is consistency among the distributions for each PaaS, but the statistics of each distribution vary based on the combination of both client and PaaS. This point is supported by the data in Table 3.

Table 3: Mean, Maximum, and Minimum Values of PaaS/Client Distributions

| Pair | Mean (s) | Min (s) | Max (s) |
| --- | --- | --- | --- |
| PaaS1/Client-A | 1397.186 | 1161.200 | 1798.600 |
| PaaS1/Client-B | 2093.313 | 1393.800 | 3896.900 |
| PaaS2/Client-A | 721.687 | 447.400 | 1089.600 |
| PaaS2/Client-B | 1045.639 | 294.600 | 1773.700 |
| PaaS3/Client-A | 289.314 | 265.700 | 348.200 |
| PaaS3/Client-B | 317.555 | 293.900 | 368.400 |

In addition, the Kolmogov-Smirnov test indicates that each of the distributions differs significantly from the others. Two conclusions may be drawn from the data. First, for any client, the expected runtime with PaaS1 is longer that that for PaaS2, which is also longer than that for PaaS3 (PaaS1 > PaaS2 > PaaS3). Thus, PaaS environments are distinct from one another in terms of performance, and this results in measurably different runtimes. The second conclusion is that for any PaaS, the expected value of the runtime differs with the client used. The results support the hypothesis that PaaS environments offer running time guarantees to hosted applications. In addition, the PaaS and client used both affect the overall runtime performance, but the results indicated that delegating the computational workload to a PaaS application diminished the impact of the client. However, even if the bulk of the computation occurs in the cloud, the client selected is important to runtime considerations. In sum, for developers and researchers, these results indicate that a PaaS application can be effectively used to handle complex computation to reduce stress upon local systems. Such a situation is practical and cost-effective because the only prerequisites for distributing computation to a PaaS are an Internet connection and an account with a PaaS provider. Each provider's environment has distinct constraints and performance, so care must be taken in choosing an environment that best suits the needs of the application to be hosted. In this study, only free services were evaluated, however each provider offers access to paid services with performance benefits. Thus, PaaS offers services that can be scaled to meet the needs of a hosted application, meaning that a high-throughput computing solution is available to anyone.

# 6 Utility As A Benchmarker

Representing a system in terms of performance metrics is the goal of any benchmarking strategy, and in the case of this experiment the representation was based on running time. The running time data was collected from three PaaSs and two clients, and there is little overlap between the dataset ranges. The shape of the data, however, shows that each PaaS/client pair has unique behavior. That the data has consistent shape despite differences in the running time ranges demonstrates that the GA captures aspects of performance unique to the system it is testing. Thus, population-based search can fill the role of benchmarker.

# 7 Future Work

This project focused on benchmarking runtimes for applications distributed This experiment demonstrated that the population-based search benchmark is useful for testing the running time performance of an application. However, running time is not the only performance measure available to gauge the quality of an application. To fully test this new benchmarking strategy, a suite needs to be created that records other performance metrics. Data collected using this suite must be judged against similar data collected using conventional techniques. In addition to the benchmarking strategy, the experiment presented in this paper may be replicated to compare performance between different types of cloud services. Investigating these benchmarks will provide more insight into any behavioral consistency available from a PaaS. In addition, a comparison of the runtime performance of local machines versus the system described in this paper could show how distributing work to a PaaS mitigates the effect of the client's own resources on running time. Finally, as most prior research has focused on the IaaS model, a comparison between identical applications run from IaaS and PaaS environments would provide a valuable benchmark of the capabilities of PaaS as a cloud service. All of this work will be valuable to the role that cloud services can provide to HPC as they continue to mature and become more accessible.

# References

[1] O. Serlin, "The history of debitcredit and the tpc." 1993.
[2] C. Binnig, D. Kossmann, T. Kraska, and S. Loesing, "How is the weather tomorrow?: Towards a benchmark for the cloud," in *Proceedings of the Second International Workshop on Testing Database Systems*, ser. DBTest '09. New York, NY, USA: ACM, 2009, pp. 9:1–9:6. [Online]. Available: http://doi.acm.org/10.1145/1594156.1594168
[3] M. Poess and C. Floyd, "New tpc benchmarks for decision support and web commerce," *ACM Sigmod Record*, vol. 29, no. 4, pp. 64–71, 2000.
[4] W. D. Smith, "Tpc-w: Benchmarking an ecommerce solution," 2000.
[5] R. Figueiredo, P. Dinda, and J. Fortes, "A case for grid computing on virtual machines," in *Distributed Computing Systems, 2003. Proceedings. 23rd International Conference on*, May 2003, pp. 550–559.
[6] W. Huang, J. Liu, B. Abali, and D. K. Panda, "A case for high performance computing with virtual machines," in *Proceedings of the 20th Annual International Conference on Supercomputing*, ser. ICS '06. New York, NY, USA: ACM, 2006, pp. 125–134. [Online]. Available: http://doi.acm.org/10.1145/1183401.1183421
[7] P. Church and A. Goscinski, "A survey of approaches and frameworks to carry out genomic data analysis on the cloud," in *Cluster, Cloud and Grid Computing (CCGrid), 2014 14th IEEE/ACM International Symposium on*, May 2014, pp. 701–710.
[8] J. Napper and P. Bientinesi, "Can cloud computing reach the top500?" in *Proceedings of the Combined Workshops on UnConventional High Performance Computing Workshop Plus Memory Access Workshop*, ser. UCHPC-MAW '09. New York, NY, USA: ACM, 2009, pp. 17–20. [Online]. Available: http://doi.acm.org/10.1145/1531666.1531671
[9] C. Vecchiola, S. Pandey, and R. Buyya, "High-performance cloud computing: A view of scientific applications," in *Proceedings of the 2009 10th International Symposium on Pervasive Systems, Algorithms, and Networks*, ser. ISPAN '09. Washington, DC, USA: IEEE Computer Society, 2009, pp. 4–16. [Online]. Available: http://dx.doi.org/10.1109/I-SPAN.2009.150
[10] W. Yu and J. S. Vetter, "Xen-based HPC: A parallel I/O perspective," in *Cluster Computing and the Grid, 2008. CCGRID '08. 8th IEEE International Symposium on*. IEEE, 2008, pp. 154–161.
[11] D. Jayasinghe, S. Malkowski, Q. Wang, J. Li, P. Xiong, and C. Pu, "Variations in performance and scalability when migrating n-tier applications to different clouds," in *Cloud Computing (CLOUD), 2011 IEEE International Conference on*, July 2011, pp. 73–80.
[12] G. Juve, E. Deelman, G. B. Berriman, B. P. Berman, and P. Maechling, "An evaluation of the cost and performance of scientific workflows on Amazon EC2," *J. Grid Comput.*, vol. 10, no. 1, pp. 5–21, Mar. 2012. [Online]. Available: http://dx.doi.org/10.1007/s10723-012-9207-6
[13] R. Tudoran, A. Costan, G. Antoniu, and L. Bougé, "A performance evaluation of Azure and Nimbus clouds for scientific applications," in *Proceedings of the 2nd International Workshop on Cloud Computing Platforms*, ser. CloudCP '12. New York, NY, USA: ACM, 2012, pp. 4:1–4:6. [Online]. Available: http://doi.acm.org/10.1145/2168697.2168701
[14] G. Jung, N. Sharma, F. Goetz, and T. Mukherjee, "Cloud capability estimation and recommendation in black-box environments using benchmark-based approximation," in *Cloud Computing (CLOUD), 2013 IEEE Sixth International Conference on*, June 2013, pp. 293–300.

[15] M. Rak and G. Aversano, "Benchmarks in the cloud: The mosaic benchmarking framework," in *Symbolic and Numeric Algorithms for Scientific Computing (SYNASC), 2012 14th International Symposium on*, Sept 2012, pp. 415–422.

[16] D. Chappell, "A short introduction to cloud platforms: An enterprise-oriented view," David Chappell and Associates, Tech. Rep., aug 2008.

[17] J. Slawinski and V. Sunderam, "Adapting MPI to MapReduce PaaS clouds: An experiment in cross-paradigm execution," in *Utility and Cloud Computing (UCC), 2012 IEEE Fifth International Conference on*, Nov 2012, pp. 199–203.

[18] A. Turner, A. Fox, J. Payne, and H. Kim, "C-mart: Benchmarking the cloud," *Parallel and Distributed Systems, IEEE Transactions on*, vol. 24, no. 6, pp. 1256–1266, June 2013.

[19] X. Bai, M. Li, B. Chen, W.-T. Tsai, and J. Gao, "Cloud testing tools," in *Service Oriented System Engineering (SOSE), 2011 IEEE 6th International Symposium on*. IEEE, 2011, pp. 1–12.

[20] D. Agarwal and S. Prasad, "Azurebench: Benchmarking the storage services of the azure cloud platform," in *Parallel and Distributed Processing Symposium Workshops PhD Forum (IPDPSW), 2012 IEEE 26th International*, May 2012, pp. 1048–1057.

[21] M. Mitchell, *An Introduction to Genetic Algorithms*, ser. A Bradford book. Bradford Books, 1998. [Online]. Available: http://books.google.com/books?id=0eznlz0TF-IC

[22] P. R. Srivastava and T.-h. Kim, "Application of genetic algorithm in software testing," *International Journal of software Engineering and its Applications*, vol. 3, no. 4, pp. 87–96, 2009.

[23] J. M. Reddy and N. D. Kumar, "Computational algorithms inspired by biological processes and evolution," *Current Science (Bangalore)*, vol. 103, no. 4, pp. 370–380, 2012.

[24] K. J. Åström and T. Hagglund, *PID Controllers: Theory, Design, and Tuning*, 2nd ed. The Instrumentation, Systems, and Automation Society, Jan. 1995.

[25] A. Swartz, "Web.py," 2013. [Online]. Available: webpy.org/